\documentclass[aps,prd,amsfonts,amsmath,nofootinbib,preprint,tightenlines,longbibliography]{revtex4-1}

\usepackage{epsf}

\def\be{\begin{equation}}
\def\ee{\end{equation}}

\begin{document}
\newcommand{\I}{\mathcal{I}}
\newcommand{\dv}{\delta v}
\newcommand{\du}{\delta u}
\newcommand{\dX}{\delta X}
\newcommand{\dd}{\text{d}}

\title{Backreaction on an infinite helical cosmic string}
\author{Ken W. Robbins, III}
\email{kenneth.robbins@tufts.edu}
\author{Ken D. Olum}
\email{kdo@cosmos.phy.tufts.edu}
\affiliation{Institute of Cosmology, Department of Physics and Astronomy,\\Tufts University, Medford, MA 02155, USA}

\begin{abstract}
To understand the properties of a possible cosmic string network
requires knowledge of the structures on long strings, which control
the breaking off of smaller loops.  These structures are influenced by
backreaction due to gravitational wave emission.  To gain an
understanding of this process, we calculate the effect of
gravitational backreaction on the ``helical breather'': an infinite
cosmic string with a helical standing wave.  We do the calculation
analytically to first order in the amplitude of the oscillation.  Our
results for the rate of loss of length agree with the total power
emitted from this system as calculated by Sakellariadou.  We also find
a rotation of the generators of the string that leads to an
advancement of the phase of the oscillation in addition to that
produced by the loss of length.
\end{abstract}
\maketitle

\section{Introduction}\label{intro}
Symmetry breaking phase transitions in the early universe may have led
to the production of topological defects, in particular cosmic
strings.  ``Cosmic superstrings'' may also have formed during the process of
string-theory-based inflation.  For a review of strings see
Ref.~\cite{Vilenkin:2000jqa} and of superstrings
\cite{Chernoff:2014cba}.





Cosmic strings release energy into gravitational radiation, leading to
potentially observable signals
\cite{Sanidas:2012tf,Sousa:2016ggw,Abbott:2017mem,Blanco-Pillado:2017oxo,Blanco-Pillado:2017rnf}.
This radiative emission reacts back on the string, decreasing its
length and changing its shape. This process affects each string on small length scales and so is
important in determining the detailed evolution of a string network
\cite{Austin:1993rg, Austin:1994dz}. Backreaction on long strings may also influence the number of loops produced and therefore the network's gravitational emission spectra \cite{Blanco-Pillado:2017oxo,Polchinski:2007rg,Chernoff:2018evo}.

The pioneering work on cosmic string backreaction was done by
Quashnock and Spergel in 1990 \cite{Quashnock:1990wv}.  More recently,
Blanco-Pillado, Olum, and Wachter have studied backreaction both
analytically
\cite{Wachter:2016hgi,Wachter:2016rwc,Blanco-Pillado:2018ael} and
numerically \cite{Blanco-Pillado:2019nto}.  All such work, however,
has studied oscillating loops of cosmic string.  Here we begin the
study of backreaction on strings that are not in the form of loops,
beginning with the simple case of the ``helical breather'', an infinite
helical standing wave whose radius oscillates between $0$ and some
value $\epsilon$.  We work in the limit where $\epsilon\ll1$, which
permits an analytic calculation of the backreaction.

In Sec.~\ref{sec:basics} we establish the general formalism that we
use to study this problem, and in Sec.~\ref{sec:breather} we discuss
the helical breather and its symmetries.  In
Sec.~\ref{sec:coordinates} we introduce a coordinate system adapted to
the backreaction problem \cite{Blanco-Pillado:2018ael}.  In
Sec.~\ref{sec:calculation} we calculate the backreaction, and in
Sec.~\ref{sec:analysis} we show how the backreaction affects the shape
of the string and check the resulting loss of energy against
the calculation of radiative power made by Sakellariadou
\cite{Sakellariadou:1990ne} for this system.  We conclude in
Sec.~\ref{sec:conclusions}.

\section{Basics}\label{sec:basics}

We follow the formalism of
Refs.~\cite{Wachter:2016rwc,Blanco-Pillado:2018ael}.  We work in the
Nambu approximation where the cosmic string is a line-like object that
traces out a two-dimensional worldsheet in spacetime.  We parameterize
the worldsheet by coordinates $(\tau,\sigma)$ where $\tau$ is temporal
and $\sigma$ is spatial, and work in the conformal gauge where the
worldsheet metric $\gamma_{ab}$ obeys $\gamma_{\tau\sigma}=0$ and
$\gamma_{\sigma\sigma}=-\gamma_{\tau\tau}$.   The motion of a string in
a flat background, before taking account of backreaction, is described
by the 4-vector
\be
X(\tau,\sigma)=\frac12\left[A(v)+B(u)\right]\,,
\ee
where we have defined null coordinates $u = \tau+\sigma$ and $v =
\tau-\sigma$, and $A$ and $B$ are 4-vector functions depending only
on $v$ and $u$ respectively, with null tangents $A'$ and $B'$.
We can further choose the worldsheet parameter $\tau$ to be the same
as the spacetime coordinate $t$, so $A'^t = B'^t = 1$.

Backreaction modifies the above evolution, by introducing acceleration
\cite{Quashnock:1990wv}
\be\label{acceleration}
X^\gamma_{,uv}=-\frac14\Gamma^\gamma_{\alpha\beta}A'^\alpha B'^\beta\,,
\ee
where $\Gamma^\gamma_{\alpha\beta}$ is the Christoffel symbol,
\be\label{eqn:Christoffel}
\Gamma^\gamma_{\alpha\beta}=\frac{1}{2}\eta^{\gamma\delta}\left(h_{\beta\delta,\alpha}+h_{\delta\alpha,\beta}-h_{\alpha\beta,\delta}\right)\,,
\ee
with $h_{\alpha\beta}$ the perturbation to the flat-space metric
$\eta_{\alpha\beta}$ due to the gravitational field of the string.

We can integrate the acceleration to find the changes to $A'$ and $B'$,
\begin{subequations}\label{eqn:delta-a'b'}\begin{align}
    \Delta A'(v) &= 2\int  du\, X_{,uv}(u,v)\,,\\
    \Delta B'(u) &= 2\int dv\, X_{,uv}(u,v)\,.
\end{align}\end{subequations} 

There is some gauge (i.e., coordinate choice) freedom in the
definition of $h_{\alpha\beta}$, that makes it difficult to separate
physical effects on the string from artifacts.  However, for a periodic
source, we can perform the integrals of Eqs.~(\ref{eqn:delta-a'b'})
over $N$ periods.  Effects that grow with $N$ are physical, while those
that oscillate may be gauge artifacts.

The metric derivatives in Eq.~(\ref{eqn:Christoffel}) can be computed
by Green's function methods.  The metric derivatives at some
observation point are found by integrating over all source points
where the string lies on the past lightcone of the observation point.
We will denote quantities at the observation point with overbars and
use the symbol $\delta$ to denote the difference between a quantity at
the source point and the same quantity at the observation point,
i.e., $\delta f= f-\bar{f}$.

\section{The helical breather}\label{sec:breather}

\begin{figure}
\centering
\epsfysize=3in\epsfbox{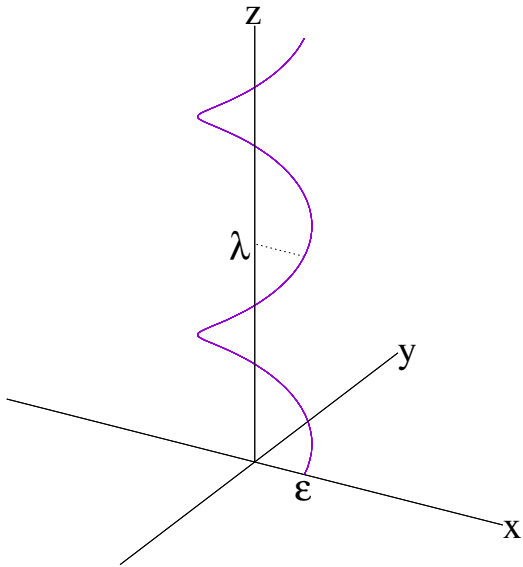}
\caption{The helical breather at $t = 0$.  The string is a constant
  distance $\epsilon$ from the $z$ axis.  The wavelength along the $z$
  direction is $\lambda=2\pi\sqrt{1-\epsilon^2}$, while the length of string
  in each period is $2\pi$.}\label{fig:helix}
\end{figure}

We will study a helical string given in Cartesian coordinates $(t, x,
y, z)$ by
\begin{subequations}\label{eqn:AB}\begin{align}
A^\alpha &= \left(v, \epsilon \cos v,-\epsilon \sin v,-v\sqrt{1-\epsilon^2}\right)\,,\\
B^\alpha &= \left(u,\epsilon\cos u,\epsilon\sin u,u\sqrt{1-\epsilon^2}\right)\,,
\end{align}\end{subequations}
so that
\be\label{eqn:X}
X^\alpha =\left(t, \epsilon \cos t \cos\sigma,\epsilon \cos t \sin\sigma,
\sigma\sqrt{1-\epsilon^2}\right)\,,
\ee
and the string is a helix winding around the $z$ axis with amplitude
varying as $\epsilon\cos t$.  See Fig.~\ref{fig:helix}.
The wavelength of the helix in $z$ is $\lambda
=2\pi\sqrt{1-\epsilon^2}$, so we will define $\lambdabar
=\sqrt{1-\epsilon^2}$, a quantity that will occur frequently.  The
energy in one winding is $2\pi\mu$, where $\mu$ is the
string tension, so the energy per unit $z$ is $\mu/\lambdabar$.

The maximum radius $\epsilon$ can run from $0$ to $1$.  Choosing
$\epsilon=0$ gives a straight string, and $\epsilon=1$ gives a
circular breather loop.  Here we will study the regime $\epsilon\ll1$
and calculate the backreaction to leading order in $\epsilon$.

The string of Eq.~(\ref{eqn:X}) has helical symmetry: it is invariant
under translation by distance $l$ in the $z$ direction combined with
counterclockwise rotation by angle $2\pi l/\lambda$.  Backreaction
will respect this symmetry, so if we evaluate Eq.~(\ref{acceleration})
at any $\sigma$ we can easily find its value at any other $\sigma$.
Different times are not equivalent, however, so we will calculate the
effects for any observation time $\bar t$ at $\bar \sigma = 0$, and
thus $\bar u =\bar  v = \bar t$.

Symmetry considerations also restrict in what directions backreaction
might act.  A rotation of the system by angle $\pi$ around around the
$x$ axis leaves the the string invariant (with the change of parameter
$\sigma\to -\sigma$).  Thus the acceleration $X^\alpha_{,uv}$
evaluated at $\bar\sigma = 0$ may have only $t$ and $x$ components, which
would be invariant under this rotation, not $y$ or $z$ components,
which would be reversed.

In fact, helical symmetry plus energy conservation almost allows us to
compute the whole evolution of the string.  To preserve helical
symmetry, the string must always be a helix with the same physical
wavelength $\lambda$.  Thus the only possibility is that the helix
shrinks toward the $z$ axis while remaining helical.  This new helix
will have smaller energy density in each winding, and the decrease of
energy density must agree with the known \cite{Sakellariadou:1990ne}
radiation rate for this system.  The only property this argument does
not determine is the phase of the oscillation after backreaction.
Doing the backreaction calculation allows to determine this phase,
which is turns out to be nontrivial, but also this calculation may
act as a starting point for similar calculations in more complex systems.
So we will calculate the effects according to
Eqs.~(\ref{acceleration}--\ref{eqn:delta-a'b'}) and check against
Ref.~\cite{Sakellariadou:1990ne} at the end.

\section{Coordinate system}\label{sec:coordinates}

Our calculations are easier in a coordinate system adapted to the
problem.  Following Ref.~\cite{Blanco-Pillado:2018ael} we define a
pseudo-orthogonal coordinate system $uvcd$, as follows. Basis vectors
$e_{(u)}=\bar{B'}/2$, $e_{(v)}=\bar{A'}/2$ are null and
their inner product is $Z/4$ where
\be
Z= \bar{A'}^\alpha\bar{B'}_\alpha= -2 (1-\epsilon^2\sin^2\bar t)\,.
\ee
The
other two basis vectors, $e_{(c)}$ and $e_{(d)}$, are spacelike unit
vectors orthogonal to $e_{(u)}$ and $e_{(v)}$ and to each other.  We
will use Greek letters from the middle of the alphabet for indices in
$uvcd$ coordinates, and Greek letters from the beginning of the
alphabet for indices in Cartesian coordinates.

In $uvcd$ coordinates, the flat metric has the form
\be
\eta_{\mu\nu}=
\begin{pmatrix}
    0 & \frac{Z}{4} & 0 & 0 \\
   \frac{Z}{4} & 0 & 0 & 0 \\
    0 & 0 & 1 & 0 \\
    0 & 0 & 0 & 1
\end{pmatrix}\,,
\ee
and Eqs.~(\ref{acceleration},\ref{eqn:Christoffel}) lead to simple expressions in $uvcd$ coordinates \cite{Blanco-Pillado:2018ael},
\begin{subequations}\label{recipe}\begin{align}
\label{recipeu}
\bar{X}^u_{,uv}&=-\frac{2}{Z}h_{vv,u}\\
\label{recipev}
\bar{X}^v_{,uv}&=-\frac{2}{Z}h_{uu,v}\\
\label{recipec}
\bar{X}^c_{,uv}&=\frac{1}{2}\left(h_{uv,c}-h_{uc,v}-h_{vc,u}\right)\\
\label{reciped}
\bar{X}^d_{,uv}&=\frac{1}{2}\left(h_{uv,d}-h_{ud,v}-h_{vd,u}\right)\,.
\end{align}\end{subequations}

We will consider observation points with $\sigma = 0$, which lie
on the $x$ axis.  Because of the symmetries of the problem, we
choose $e_{(c)}$ to have only $y$ and $z$ components, while $e_{(d)}$
has only $t$ and $x$ components.  This gives
\begin{subequations}\begin{align}
e^\alpha_{(u)}& =\left(\frac{1}{2},-\frac{\epsilon}{2}\sin \bar{t},\frac{\epsilon}{2}\cos \bar{t},\frac{\lambdabar}{2}\right)\\
e^\alpha_{(v)}& =\left(\frac{1}{2},-\frac{\epsilon}{2}\sin \bar{t},-\frac{\epsilon}{2}\cos \bar{t},-\frac{\lambdabar}{2}\right)\\
e^\alpha_{(c)}& =\left(0,0,-\lambdabar\gamma,\epsilon\gamma\cos \bar{t}\right)\\
\label{eqn:ed}
e^\alpha_{(d)}& =\left(-\epsilon\gamma\sin \bar{t},\gamma,0,0\right)\,,
\end{align}\end{subequations}
where $\gamma =\sqrt{-2/Z} = 1/\sqrt{1-\epsilon^2\sin^2\bar t}$.  This
is the Lorentz factor associated with the string velocity,
$-\epsilon\sin\bar t$.  In the particular case of $\bar{t}=0$, when
the string is momentarily stationary, $e_{(d)}=e_{(x)}$
corresponds to the radial direction of the helix.  At other times,
$e_{(d)}$ is just a boosted version of that radial vector.

We hope there will be no confusion arising from our use of $u,v$ as
labels for both the worldsheet and the spacetime coordinates.  The
coordinate systems are closely related: at the observation point, the
motion generated by changing the worldsheet coordinate $u$ is just the
spacetime vector $e_{(u)}$ and the same for $v$.  The metric
$h_{\mu\nu}$ is a function of spacetime position, so the notations
$h_{\mu\nu,u}$ and $h_{\mu\nu,v}$ mean derivatives with respect to the
$u$ and $v$ directions in spacetime.  In all other cases, subscripts
``$,u$'' and ``$,v$'' mean derivatives with respect to the worldsheet
variables.  All components of vectors and tensors are spacetime
components.

We can convert the components of a contravariant vector from $uvcd$
coordinates to Cartesian via $V^\alpha = {M^\alpha}_\mu V^\mu$, where
\be
{M^\alpha}_\mu =\begin{pmatrix} e_{(u)} & \vrule & e_{(v)}& \vrule &
e_{(c)}& \vrule & e_{(d)}\end{pmatrix}\,.
\ee
The transpose of this matrix will convert covariant components from
Cartesian to $uvcd$, $V_\mu = {(M^T)_\mu}^\alpha V_\alpha$.  In
computing the metric derivatives we will often want to transform
contravariant vectors to the $uvcd$ coordinates and lower the index,
which we can do by $V_\mu = {(M^T)_\mu}^\alpha \eta_{\alpha\beta}
V^\beta$.  The components of the transformation matrix are
\be\label{cartu}
{(M^T)_\mu}^\alpha \eta_{\alpha\beta} =\begin{pmatrix}
   -\frac{1}{2} & -\frac{\epsilon}{2}\sin \bar{t} & \frac{\epsilon}{2}\cos \bar{t} & \frac{\lambdabar}{2} \\
   -\frac{1}{2} & -\frac{\epsilon}{2}\sin \bar{t} & -\frac{\epsilon}{2}\cos \bar{t} & -\frac{\lambdabar}{2} \\
    0 & 0 &-\lambdabar\gamma &\epsilon\gamma\cos \bar{t} \\
    \epsilon\gamma\sin \bar{t} & \gamma & 0 & 0
\end{pmatrix}\,.
\ee

\section{Calculation}\label{sec:calculation}

We can calculate the backreaction at the observation point $\bar X$ by
integrating over all possible source points on the string world sheet
that lie on the backward lightcone of $\bar X$.  The vector pointing
from $\bar X$ to any worldsheet point $X(u, v)$ is
\be\label{cartx}
\dX^\alpha=\frac{1}{2}(\du+\dv,\epsilon\left(\omega_c\left(\du, \bar{t}\right)+\omega_c\left(\dv, \bar{t}\right)\right),\epsilon\left(\omega_s\left(\du, \bar{t}\right)-\omega_s\left(\dv, \bar{t}\right)\right),\lambdabar\left(\du-\dv\right))\,,
\ee
where we have defined
\begin{subequations}\begin{align}
\omega_s(a,b)&=\sin\left(a+b\right)-\sin b\\
\omega_c(a, b)&=\cos\left(a+b\right)-\cos b\,.
\end{align}\end{subequations}
The condition to be on the lightcone is that the interval $\I=\delta
X^\alpha \dX_\alpha= 0$.  We can write
\be\label{null} 
\I=-\du\dv
+\frac{\epsilon^2}{4}\left[\left(\omega_c\left(\du,
\bar{t}\right)+\omega_c\left(\dv,
\bar{t}\right)\right)^2+\left(\omega_s\left(\du,
\bar{t}\right)-\omega_s\left(\dv,
\bar{t}\right)\right)^2-\left(\du-\dv\right)^2\right]\,.
\ee
The intersection of the worldsheet has two branches, one that starts
out in the negative $u$ direction and the other that starts out in the
negative $v$ direction~\cite{Blanco-Pillado:2019nto}.  We will choose
the former branch and add in the latter branch by symmetry.
Considerations above assure us that radial and temporal contributions to the
backreaction from the other branch are equal and constructive, while
other spatial directions are equal and destructive.

We thus define $\dv(\du)$ to be the $\dv$ that makes $\I = 0$ for the
given $\du$.  However, the transcendental nature of Eq.~(\ref{null})
makes it impossible to write $\dv(\du)$ in closed form, so we must
approximate.  From this point forward we will take $\epsilon\ll1$, and
keep only terms that contribute at leading order in $\epsilon$.
With $\epsilon = 0$, $\I = 0$ would give $\dv = 0$.  Using this on the
right hand side of Eq.~(\ref{null}), we find
\be\label{vu}
\dv(\du)=-\frac{\epsilon^2\du}{4}+\frac{\epsilon^2}{\du}\sin^2\frac{\du}{2}\,.
\ee
In most cases only the first term on the right hand side will contribute.

The metric derivatives can be found by integrating along our chosen
branch of the backward lightcone \cite{Quashnock:1990wv},
\be \label{mder}
h_{\mu\nu,\lambda}=4G\mu\int^0_{-\infty}\dd\du\,\I_{,v}^{-1}\left[\frac{\partial}{\partial v}\left(s_{\mu\nu} \dX_\lambda\I_{,v}^{-1}\right)\right]_{\dv= \dv(\du)}
\ee
where $G$ is Newton's constant, $s_{\mu\nu}=\Sigma_{\mu\nu}\left(A', B'\right)$, with
\be\label{eqn:Sigma}
\Sigma_{\mu\nu}\left(P,Q\right)= P_\mu Q_\nu+P_\nu Q_\mu-\eta_{\mu\nu}\left(P^\lambda Q_\lambda\right)\,,
\ee
and $h_{\mu\nu,\lambda}$ here denotes only the contribution from
the one branch.  There is a cancellation in the $uv$ term in Eq.~(\ref{eqn:Sigma})
\cite{Blanco-Pillado:2019nto},
\be
\Sigma_{uv}\left(P,Q\right)=-\frac{Z}{4}(P_cQ_c+P_dQ_d)\,.
\ee

We then calculate $\I_{,v}$ and the the necessary components of
$s_{\mu\nu}$ and $\dX_\mu$,
\begin{subequations}\label{eqn:components}\begin{align}
s_{uu}&=\epsilon^2(1-\cos\du)\\
s_{vv}&=\epsilon^2(1-\cos\dv)\\
s_{uv}&=\frac{\epsilon^2}{2}\left[\omega_s\left(\du,
\bar{t}\right)\omega_s\left(\dv, \bar{t}\right)
-\omega_c\left(\du,\bar{t}\right)\omega_c\left(\dv, \bar{t}\right)\right]\\
s_{ud}&=\left[\epsilon+\frac{\epsilon^3}{2}\sin^2\bar{t}-\epsilon^3\sin^2\frac{\dv+2\bar{t}}{2}\right]\omega_s\left(\du, \bar{t}\right)+\epsilon^3\sin^2\frac{\du}{2}\omega_s\left(\dv, \bar{t}\right)\\
s_{vd}&=\left[\epsilon+\frac{\epsilon^3}{2}\sin^2\bar{t}-\epsilon^3\sin^2\frac{\du+2\bar{t}}{2}\right]\omega_s\left(\dv, \bar{t}\right)+\epsilon^3\sin^2\frac{\dv}{2}\omega_s\left(\du, \bar{t}\right)
\end{align}\end{subequations}
\begin{subequations}\begin{align}
\dX_u&=-\frac{\dv}{2}+\frac{\epsilon^2}{4}\left[\dv-\du+\left(\omega_s\left(\du, \bar{t}\right)-\omega_s\left(\dv, \bar{t}\right)\right)\cos \bar{t}-\left(\omega_c\left(\du, \bar{t}\right)+\omega_c\left(\dv, \bar{t}\right)\right)\sin \bar{t}\right]\nonumber\\
&=-\frac{\dv}{2}+\frac{\epsilon^2}{4}\left[\dv-\du+
\sin\du -\omega_s(\dv, 2\bar t)\right]\\
\dX_v&=
-\frac{\du}{2}+\frac{\epsilon^2}{4}\left[\du-\dv+\left(\omega_s\left(\dv, \bar{t}\right)-\omega_s\left(\du, \bar{t}\right)\right)\cos \bar{t}-\left(\omega_c\left(\du, \bar{t}\right)+\omega_c\left(\dv, \bar{t}\right)\right)\sin \bar{t}\right]\nonumber\\
&=-\frac{\du}{2}+\frac{\epsilon^2}{4}\left[\du-\dv+
\sin\dv -\omega_s(\du, 2\bar t)\right]\\
\label{eqn:Xd}\dX_d&=\frac{\epsilon}{2}\left[\left(\du+\dv\right)\sin
\bar{t}+\omega_c\left(\dv, \bar{t}\right)+\omega_c\left(\du, \bar{t}\right)\right]
\end{align}\end{subequations}
\be\label{eqn:in2}
\I_{,v}=-\du+\frac{\epsilon^2}{2}\left[\du-\dv
+\sin\dv-\omega_s\left(\du,\dv+2\bar{t}\right)\right]\,.
\ee
Here are the orders in $\epsilon$ of the above quantities and their derivatives.
\begin{subequations}\label{eqn:powers}\begin{align}
s_{uu}&\sim\epsilon^2 &s_{uu,v}&=0\label{suupower} \\
s_{vv}&\sim\epsilon^2 &s_{vv,v}&\sim\epsilon^2\label{svvpower}\\
s_{uv}&\sim\epsilon^2 &s_{uv,v}&\sim\epsilon^2\label{suvpower}\\
s_{ud}&\sim\epsilon &s_{ud,v}&\sim\epsilon^3\label{sudpower}\\
s_{vd}&\sim\epsilon & s_{vd,v}&\sim\epsilon \label{svdpower}\\
\dX_u&\sim\dv+\epsilon^2 &\dX_{u,v}&\sim1+\epsilon^2\label{Xupower}\\
\dX_v&\sim\du+\epsilon^2 &\dX_{v,v}&\sim\epsilon^2\label{Xvpower} \\
\dX_d&\sim\epsilon &\dX_{d,v}&\sim\epsilon\label{Xdpower}\\
\I_{,v}&\sim\du+\epsilon^2 &\I_{,vv}&\sim\epsilon^2\label{Ivpower}\,.
\end{align}\end{subequations}
We can use Eqs.~(\ref{eqn:powers}) to see which terms we need to keep
in our calculation.

Equations~(\ref{eqn:components}--\ref{eqn:powers}) apply to all $\du$
and $\dv$.  After we take the derivative and set $\dv$ according to Equation~(\ref{vu}) we can make stronger statements.  First,
terms in Eqs.~(\ref{eqn:components}--\ref{eqn:in2}) that are simply proportional to
$\dv$ have 2 additional powers of $\epsilon$.  When $\dv$ appears
in the argument of a trigonometric function, the situation is somewhat
more complicated.  Consider an integral of the form
\be\label{eqn:dvint}
\int^0_{-\infty}\dd\du\frac{\omega(\dv,v_1)}{\du^2}f(\du)\,,
\ee
where $\omega$ is $\omega_s$ or $\omega_c$, $v_1$ is arbitrary, and
$|f(\du)|$ is bounded by some constant.  This integral is small
because $\omega(\dv,v_1)\ll1$ until $\du\sim\epsilon^{-2}$, but then the
integrand is suppressed by the large denominator of order
$\epsilon^{-4}$.  Specifically, both $\omega_c$ and $\omega_s$ obey
\be\label{eqn:omegabound}
|\omega(\dv,v_1)|< 2
\ee
and also
\be\label{eqn:omegavbound}
|\omega(\dv,\bar t)|< \dv\,,
\ee
since $|\dd\omega(\dv,\bar t)/\dd\dv| < 1$.
Now we break up the integral at $-\epsilon^{-2}$.  Using
Eq.~(\ref{eqn:omegabound}),
\be
\left|\int^{-\epsilon^{-2}}_{-\infty}\dd\du\frac{\omega(\dv,\bar t)}{\du^2}\right|
< 2\epsilon^2\,,
\ee
and using Eq.~(\ref{eqn:omegavbound}), 
\be
\left|\int^0_{\epsilon^{-2}}\dd\du\frac{\omega(\dv,\bar t)}{\du^2}\right|
< \left|\int^0_{-\epsilon^{-2}}\dd\du\frac{\dv}{\du^2}\right|\,.
\ee
As $\du\to 0$, $\dv\sim\du^3$, so there is no divergence near that
limit.  For $|\du|\gg 1$, $\dv\sim\ -\epsilon^2\du/4$, giving a
logarithmic divergence\footnote{The integral can also be done in
  closed form.}, so
\be
\int_0^{\epsilon^{-2}}\dd\du\frac{\dv}{\du^2}\sim \epsilon^2\ln\epsilon\,,
\ee
and since $f$ is bounded by a constant,
\be
\int^0_{-\infty}\dd\du\frac{\omega(\dv,v_1)}{\du^2}f(\du)\sim \epsilon^2\ln\epsilon
\ee
at most.

This argument does not hold if $\du^2$ is replaced by $\du$ in the
denominator of Eq.~(\ref{eqn:dvint}).  For example
\be
\int_0^\infty\dd \du \frac{\sin\dv}{\du} =\frac{\pi}{2}+O(\epsilon^2)\,.
\ee

Now we calculate the components of the backreaction using
Eqs.~(\ref{recipe}, \ref{mder}).  We will compute only the branch that
starts in the negative $\du$ direction and add the other by symmetry.
This doubles $\bar{X}^d_{,uv}$, cancels $\bar{X}^c_{,uv}$ so that we
don't have to compute it, and adds our computed $\bar{X}^u_{,uv}$ and
$\bar{X}^v_{,uv}$ to give a common backreaction in these two
directions.  These combinations guarantee that backreaction at
$\bar\sigma = 0$ acts only in the $t$ and $x$ directions, as discussed
above.

We will find $\bar{X}^d_{,uv}\sim \epsilon^3$ and $\bar{X}^v_{,uv}\sim
\epsilon^4$, so we will ignore terms of higher orders.

\subsection{$h_{uv,d}$}
We start by computing $\bar{X}^d_{,uv}$, following
Eq.~(\ref{reciped}).  The first metric derivative, $h_{uv,d}$, is the
simplest. First, using
Equations~(\ref{suvpower},\ref{Xdpower},\ref{Ivpower}) we see that the
term with $\I_{,vv}$ is negligible.  Then, when we differentiate
$\dX_d$ we get
\be
s_{uv}\dX_{d,v}\I_{,v}^{-2} =
\frac{\epsilon}{2}\int^0_{-\infty}\dd\du\frac{s_{uv}\omega_s(\dv,\bar t)}{\du^2}\,.
\ee
Since $|s_{uv}|<\epsilon^2$, this has the form of
Eq.~(\ref{eqn:dvint}) with a prefactor proportional to $\epsilon^3$,
so it does not contribute.

The remaining term depends on $s_{uv,v} =
(\epsilon/2)\omega_s(\du,\dv+2\bar{t})$ and $\delta X_d$ from
Eq.~(\ref{eqn:Xd}).  The term involving $\dv\sin\bar t$ is immediately
of order $\epsilon^5$.  The term involving $\omega_c(\dv,\bar t)$ has
the form of Eq.~(\ref{eqn:dvint}) and gives $\epsilon^5\ln\epsilon$
at most.  Ignoring these leaves us with
\be\label{huvd}
h_{uv,d}=\int^0_{-\infty}\dd\du\,\frac{G\mu\epsilon^3}{\du^2}\omega_s\left(\du,\dv+2\bar{t}\right)\left(\omega_c\left(\du,\bar{t}\right)+\du\sin \bar{t}\right)\,.
\ee 

\subsection{$h_{ud,v}$}

For the second metric derivative $h_{ud,v}$ we use
Eqs.~(\ref{sudpower}, \ref{Xvpower}, \ref{Ivpower}) and see that we
can differentiate any one of the factors and still have something of
order $\epsilon^3$ in all.  But closer inspection reveals that the
$\dX_{v,v}$ term goes as $\epsilon^2\left(1-\cos\dv\right)$, which
leads to an integral in the form of Eq.~(\ref{eqn:dvint}) that does
not contribute.

For the other terms we can use $\dX_v = -\du/2$.  We then
differentiate one of $s_{ud,v}$ and $\I_{,v}$ and expand both to
leading order in $\epsilon$,
\begin{align}
s_{ud,v} &=
\frac{\epsilon^3}{2}\sin\left(\dv+2\bar{t}\right)\omega_s\left(\du,\bar{t}\right) 
+\epsilon^3 \sin^2\frac{\du}{2}\cos(\dv+\bar{t})
& \I_{,v} &= -\du\\
\I_{,vv} &=
\frac{\epsilon^2}{2}\left(\cos\dv-1-\omega_c(\du,\dv+2\bar t)\right)
& s_{ud} & = \epsilon\omega_s(\du,\bar t)
\end{align}

Thus we find
\be
\begin{split}\label{hudv}
h_{ud,v}=\int^0_{-\infty}\dd\du\,\frac{G\mu\epsilon^3}{\du^2}\bigg(&\du\sin\left(\dv+2\bar{t}\right)\omega_s\left(\du,\bar{t}\right)-2\du \sin^2\frac{\du}{2}\cos\left(\dv+\bar{t}\right)\\
&+\omega_s\left(\du,\bar{t}\right)\omega_c\left(\du,\dv+2\bar{t}\right)\bigg)
\end{split}
\ee
plus a term of the form of Eq.~(\ref{eqn:dvint}) that does not
contribute.

\subsection{$h_{vd,u}$}
The final metric derivative for $\bar{X}^d_{,uv}$ uses
Eqs~(\ref{svdpower},\ref{Xupower},\ref{Ivpower}).  If we don't
differentiate $\delta X_u$, it goes as $\epsilon^2$.  Thus the term
involving $\I_{,vv}$ goes as $\epsilon^5$.  When we differentiate
$s_{vd}$ we find
\be\label{eqn:hvdu1}
h_{vd,u}\supseteq -G\mu\epsilon\int_{-\infty}^0\frac{\dd\du}{\du^2}
\cos(\dv+\bar t)\left(2\dv+\epsilon^2(\du-\sin\du+\omega_s(\dv, 2\bar t))\right)\,.
\ee
When we differentiate $\delta X_u$, the leading order effect is
\be\label{eqn:hvdu2}
h_{vd,u}\supseteq -2G\mu\epsilon\int_{-\infty}^0\frac{\dd\du}{\du^2}
\omega_s(\dv, \bar t)\,.
\ee
This is superficially $\sim\epsilon$, however the argument of
Eq.~(\ref{eqn:dvint}) shows that it contributes only at order
$\epsilon^3$.  In principle there could be a contribution from expanding
any of $s_{vd}$, $\delta X_{u,v}$, and $\I_{,v}$ to next order in
$\epsilon$, but one can check that no such term contributes at order
$\epsilon^3$.

Integrating Eq.~(\ref{eqn:hvdu2}) by parts gives
\be\label{eqn:hvdu3}
-2G\mu\epsilon\int_{-\infty}^0\frac{\dd\du}{\du}
\frac{\dd}{\dd\du}\omega_s(\dv, \bar t)
= -2G\mu\epsilon\int_{-\infty}^0\frac{\dd\du}{\du}
\cos(\dv+\bar t)\frac{\dd\dv}{\dd\du}\,.
\ee
From Eq.~(\ref{vu}) we find
\be
\frac{\dd\dv}{\dd\du}= -\frac{\dv}{\du} -\frac{\epsilon^2}{2}\du
+\frac{\epsilon^2}{2\du}\sin\du\,.
\ee
When we add Eqs.~(\ref{eqn:hvdu1},\ref{eqn:hvdu3}), all terms cancel
except the term with $\omega_s$, which does not contribute because it
has the form of Eq.~(\ref{eqn:dvint}).  Thus $h_{vd,u}=0$.

\subsection{$\bar{X}^d_{,uv}$}

To compute $\bar{X}^d_{,uv}$ we combine Eqs.~(\ref{huvd},\ref{hudv})
according to Eq.~(\ref{reciped}).  All metric derivatives above are
computed only on one of the two branches.  Adding the other branch
multiplies the result by 2, so we just subtract Eq.~(\ref{hudv}) from
Eq.~(\ref{huvd}).  Many terms cancel, and we find
\be\label{midthing}
\bar{X}^d_{,uv}=G\mu\epsilon^3
\int^0_{-\infty}\frac{\dd\du}{\du}
\left[\cos(\dv+\bar{t})-\cos(\dv-\du+\bar{t})
+\frac{2}{\du}\sin(\dv+\bar{t})\left(1-\cos\du\right)\right]\,.
\ee
The overall result is
\be\label{finale}
\bar{X}^d_{,uv}=G\mu\epsilon^3\left[\cos \bar{t}\ln\frac{\epsilon^2}{4}+\pi\sin \bar{t}\right]\,.
\ee

\subsection{$h_{vv,u}$}
Equation~(\ref{finale}) will contribute a term proportional to
$\epsilon^4$ to the time component of $\bar{X}_{,uv}$.  Such a term
might also arise from an $\epsilon^4$ in $\bar{X}^v_{,uv}$ and
$\bar{X}^u_{,uv}$, so we now compute those terms at that order.

The only metric derivative appearing in $\bar{X}^u_{,uv}$ is
$h_{vv,u}$. From Eqs~(\ref{svvpower},\ref{Xupower},\ref{Ivpower}), the
term where we differentiate $\I_{,v}$ will go as $\epsilon^6$ and can
be ignored.  The term where we differentiate $s_{vv}$ contributes
\be
h_{vv,u}\supseteq -2G\mu\int^0_{-\infty}\dd\du \frac{\epsilon^2\dv}{\du^2}\sin\dv\,.
\ee
We replace the $\dv$ that stands outside the sine function, keeping only the
first term of Eq.~(\ref{vu}), because the second term contributes at
higher order in $\epsilon$ by the same analysis as
Eq.~(\ref{eqn:dvint}), giving
\be\label{eqn:hvvu1}
h_{vv,u}\supseteq G\mu\int^0_{-\infty}\dd\du \frac{\epsilon^4}{2\du}\sin\dv\,.
\ee
Finally, when we differentiate $\delta X_u$, we get a term
proportional to $\epsilon^2$,
\be\label{eqn:hvvu2}
h_{vv,u}\supseteq -G\mu\int^0_{-\infty}\dd\du \frac{2\epsilon^2}{\du^2}(1-\cos\dv)\,.
\ee
We should also expand $\delta X_{u,v}$ and $I_{,v}^{-2}$ to next order
in $\epsilon^2$, but all such terms vanish because of Eq.~(\ref{eqn:dvint}).

When we combine Eqs.~(\ref{eqn:hvvu1},\ref{eqn:hvvu2}), the integrand
is just the derivative with respect to $\delta u$ of $2\epsilon^2
(1-\cos\dv)/\du$, which vanishes at the limits of integration, plus
terms in the form of Eq.~(\ref{eqn:dvint}) which are negligible. Thus
the integral evaluates to zero and $\bar{X}^u_{,uv}=0$.

\subsection{$h_{uu,v}$}
For $\bar{X}^v_{,uv}$ we need to compute $h_{uu,v}$, following
Eqs~(\ref{suupower},\ref{Xvpower},\ref{Ivpower}). First notice that
$s_{uu}$ doesn't depend on $\dv$, so when the derivative acts on
it we get no contribution.  The derivatives of $\I_{,v}$ and $\delta
X_v$ each go as $\epsilon^2$, as does $s_{uu}$.  So the result goes as
$\epsilon^4$ and we can ignore subleading terms in $\I_{,v}$ and
$\delta X_v$.  When we differentiate $\delta X_v$, the result has the
form of Eq.~(\ref{eqn:dvint}) and does not contribute.  When we
differentiate $\I_{,v}$ and ignore terms that do not contribute
according to Eq.~(\ref{eqn:dvint}), we are left with
\begin{equation}\label{vbackreact}
h_{uu,v}=G\mu\int^0_{-\infty}\dd\du\frac{\epsilon^4}{\du^2}(1-\cos\du)\omega_c\left(\du,2\bar{t}\right)=G \mu \frac{\epsilon^4}{2}\left(\sin2\bar{t}\ln4-\pi\cos2\bar{t}\right)\,.
\end{equation}
\subsection{$X^u_{,uv}$ and $X^v_{,uv}$}
We now compute $X^u_{,uv}$ and $X^v_{,uv}$ following
Eq.~(\ref{recipev}).  Since the contribution already goes as
$\epsilon^4$, we can ignore the factor $-2/Z= 1+O(\epsilon^2)$.  Equation~(\ref{vbackreact})
gives the contribution from the branch that starts in the negative
$\du$ direction.  The other branch gives the same quantity in
$h_{vv,u}$ and nothing in $h_{uu,v}$ so we find in all
\be\label{uvbackreact}
\bar{X}^v_{,uv} = \bar{X}^u_{,uv} =
G\mu\frac{\epsilon^4}{2}\left(\sin2\bar{t}\ln4-\pi\cos2\bar{t}\right)\,.
\ee

\section{Analysis}\label{sec:analysis}

\subsection{Effect on tangent vectors}

Now that we have $\bar{X}_{,uv}$, let us compute its effect on the
tangent vectors $A'$ and $B'$.  We will now drop the overbars because
we are concerned only with the observation point.
References~\cite{Quashnock:1990wv,Blanco-Pillado:2019nto} integrated
$X_{,uv}$ around the loop to find the total change to the tangent
vectors.  Here we don't have a loop, but we do have a periodic system,
with period $2\pi$.  So we can start with the unperturbed vector,
\be\label{eqn:A0}
A'^\alpha= (1,0,-\epsilon,-\sqrt{1-\epsilon^2})
\ee
at $v = 0$ and write the change in one oscillation
as\footnote{This integral goes to $4\pi$ because $t =
  (u+v)/2$ needs to reach $2\pi$ and we are keeping $v = 0$.  The
  corresponding situation for a loop is that $A$ and $B$ have period $L$
  but the loop oscillates with period $L/2$.}
\be\label{eqn:dA0}
\Delta A' = 2\int^{4\pi}_0 du\, X_{,uv}(u,0)\,.
\ee
Let us first first consider the effect of $X_{,uv}^d$ given by
Eq.~(\ref{finale}).  This multiplies $e_{(d)}$, but we must
generalize the form given in Eq.~(\ref{eqn:ed}) to account for the
fact that we no longer have $\sigma = 0$, but rather $\sigma = t =
u/2$.  This produces an overall rotation of the coordinate system by
angle $\sigma$, so
\be\label{eqn:edrot}
e^\alpha_{(d)} =\left (-\epsilon\sin(u/2),\cos(u/2),\sin(u/2),0\right)\,,
\ee
where we have dropped factors of order $\epsilon^2$.  

Now multiply Eq.~(\ref{eqn:edrot}) by Eq.~(\ref{finale}).  Using
$\int_0^{4\pi}\dd u\sin^2(u/2) =\int_0^{4\pi}\dd u\cos^2(u/2)=2\pi$, 
$\int_0^{4\pi}\dd u\sin(u/2)\cos(u/2)= 0$, we find
\be\label{eqn:dA}
\Delta A'^\alpha = 4\pi G\mu\epsilon^3
\left(-\pi\epsilon,\ln\frac{\epsilon^2}{4},\pi,0\right)\,.
\ee

Equation~(\ref{eqn:dA}) is in fact the entire backreaction, because
the remaining components, $X^u_{,uv}$ and $X^v_{,uv}$ do not
contribute to $\Delta A'$.  When we multiply them by $e_{(u)}$ and
$e_{(v)}$ respectively, the Cartesian $y$ and $z$ components cancel,
and the $x$ component is suppressed by an additional power of
$\epsilon$.  That leaves the time component, but it oscillates with
angular frequency 2 and vanishes on integration.\footnote{Following
  Ref.~\cite{Wachter:2016rwc}, our strategy for distinguishing gauge
  artifacts from physical effects is that the former vanish on
  integration, while the latter accumulate.  Thus we do not know
  whether those terms that vanish on integration have any physical
  significance.}

The change to the time component in Eq.~(\ref{eqn:dA}) is just the
fractional change in the energy of the string in one oscillation
\cite{Blanco-Pillado:2019nto}, in this case $-4\pi^2 G\mu\epsilon^4$.
Multiplying by the energy per unit length $\mu$ and dividing by the
period gives the average rate of change of energy per unit length,
$-2\pi G\mu^2\epsilon^4$, in agreement with the radiation rate
computed by Sakellariadou~\cite{Sakellariadou:1990ne} in the limit
$\epsilon\ll1$.

To find the perturbed $A'$, we add Eq.~(\ref{eqn:dA0}) to
Eq.~(\ref{eqn:A0}) but then rescale to return the time component of
$A'$ to 1,
\begin{align}\label{eqn:newA}
A'^\alpha_{(1)} &= \left(1,4\pi G\mu\epsilon^3\ln\frac{\epsilon^2}{4},
-(\epsilon -4\pi^2 G\mu\epsilon^3), -\frac{\sqrt{1-\epsilon^2}}{1-4\pi^2 G\mu\epsilon^4}\right)\\
&= \left(1,4\pi G\mu\epsilon^3\ln\frac{\epsilon^2}{4},
-\epsilon_{(1)}, -\sqrt{1-{\epsilon_{(1)}}^2}\right)
\end{align}
with $\epsilon_{(1)}=\epsilon(1-4\pi^2 G\mu\epsilon^2)$.  We ignored
terms of order $\epsilon^6$ and higher.
By symmetry, $B'$ will be modified to
\be
\label{eqn:newB}
B'^\alpha_{(1)} = \left(1,4\pi G\mu\epsilon^3\ln\frac{\epsilon^2}{4},
\epsilon_{(1)}, \sqrt{1-{\epsilon_{(1)}}^2}\right)\,.
\ee
Thus the
helix now has amplitude $\epsilon_{(1)}$, and thus the energy per unit
distance in $z$ is reduced to $\mu(1+\epsilon_{(1)}^2/2)$, lower by 
$4\pi^2 G\mu^2\epsilon^4$ than before in agreement with the above.

In addition, the new $x$ components in
Eq.~(\ref{eqn:newA},\ref{eqn:newB}) represent a rotation of each
vector through angle $4\pi G\mu\epsilon^2\ln(\epsilon^2/4)$.  This
rotation obeys the helical symmetry of the string, but it advances the
time that the string comes to rest (i.e., when $A'$ and $B'$
point in spatially opposite directions) in successive oscillations.
In the absence of any backreaction, the string would come to rest at
its original position at times $2\pi N$ where $N$ is an integer.  The
effect of the rotation is to advance these times by
\be\label{eqn:dtrot}
\Delta T_N^{(\text{rot})} = 4\pi N G\mu\epsilon^2\ln(\epsilon^2/4)\,.
\ee
On the other hand, the loss of energy decreases the oscillation period
so that the period after $N$ oscillations has decreased by $8\pi^3 N
G\mu\epsilon^4$, and thus the end of the $N$th oscillation is offset
by time
\be\label{eqn:dtrad}
\Delta T_N^{(\text{rad})} \approx -4\pi^3 N^2 G\mu\epsilon^4\,.
\ee
Since $\Delta T_N^{(\text{rad})}$ grows quadratically with $N$, it
will eventually be the dominant effect controlling the times at which
the helix reaches its maximum size.  However, $\Delta
T_N^{(\text{rot})}$ starts much larger, because it has two fewer powers
of $\epsilon$.  There is also an additional logarithmic term.  Thus if
one were to make a precise observation of a small-$\epsilon$ helix,
the first effect one would notice is $\Delta T_N^{(\text{rot})}$.

\subsection{Analysis of power with backreaction treated as friction}

As a simple check, we can treat the acceleration due to backreaction
as arising from a force akin to frictional damping and compute the
average power lost to that force.  The effective force per unit length
is the acceleration times the linear energy density $\mu$.  It acts in
the $x$ direction with magnitude $\mu X^x_{,tt}$.

Since $(u,v)=(\tau+\sigma, \tau-\sigma)$, $\partial/\partial
u=\left(\partial/\partial\tau+\partial/\partial\sigma\right)/2$ and
$\partial/\partial
v=\left(\partial/\partial\tau-\partial/\partial\sigma\right)/2$, so
$\partial^2/\partial u\partial
v=\left(\partial^2/\partial\tau^2-\partial^2/\partial\sigma^2\right)/4$.
Without backreaction the quantity on the right would vanish, and the
effect of backreaction is to introduce an additional acceleration
$X_{,tt} =4X_{,uv}$.  In the $x$ direction,
\be
X^x_{,tt} = 4G\mu\epsilon^3\left(\cos t\ln\frac{\epsilon^2}{4}+\pi\sin t \right)\,,
\ee
and the velocity in
this direction is $X^d_{,t}=-\epsilon\sin t$.  To find the
rate of change of energy per unit length we multiply the frictional force by the
velocity and average over one period,
\be\label{rawpower}
-\frac{2G\mu^2\epsilon^4}{\pi}\int^{2\pi}_0\dd t\,\sin t\left(\cos
t\ln\frac{\epsilon^2}{4}+\pi\sin t\right)=-2\pi G\mu^2\epsilon^4\,.
\ee
again in agreement with the calculation of power radiated in
Ref.~\cite{Sakellariadou:1990ne}.

\section{Conclusion}\label{sec:conclusions}

We calculated the effective backreaction on the helical breather in
the limit of small amplitude $\epsilon$.  We found a secular decrease
of length in agreement with previous work \cite{Sakellariadou:1990ne}
for the power radiated.  We also found a rotation of the $A'$ and $B'$
vectors, which dominates the behavior at early times.  This is not
associated with loss of energy to gravitational radiation, so one may
say that it is a gravitational effect of the string on itself rather
than the backreaction of radiation emission.

In addition to the two effects above, our calculation showed some
effects that vanish when one integrates over time.  It is not clear
whether these should be thought of as conservative forces that take
energy from the string for part of the oscillation period and return
it at others, or just considered gauge artifacts.  To resolve this
question would require a clear definition of what one means by
acceleration of the trajectory of the string as it moves in a
time-varying spacetime.

In the long run we would like to understand the effect of backreaction
on long strings in a cosmological network.  Does backreaction smooth
the string on a certain scale, so that no loops are produced below
that scale?  This question will have to await future work, but we hope
that the present analysis of a situation simple enough to be
calculated exactly (in the small-amplitude limit) will act as a
starting point for further investigation.

\section{Acknowledgments}

We thank J. J. Blanco-Pillado and Jeremy Wachter for helpful
conversations.  This work was supported in part by the National
Science Foundation under Grants No.~1520792 and No.~1820902.

\bibliography{postrefbib1}

\end{document}